# Magnetic nanoparticles with bulk-like properties

Xavier Batlle[a)], N. Pérez, P. Guardia, O. Iglesias and A. Labarta

*Dep. Física Fonamental and Institut de Nanociència i Nanotecnologia (IN2UB), Universitat*

*de Barcelona, Martí i Franquès 1, 08028 Barcelona, Spain*

F. Bartolomé, L.M. García and J. Bartolomé

*Instituto de Ciencia de Materiales de Aragón (ICMA)-CSIC and Dep. Física Materia*

*Condensada, Universidad de Zaragoza, 50009 Zaragoza, Spain*

A.G. Roca, M. P. Morales and C.J. Serna

*Instituto de Ciencia de Materiales de Madrid (ICMM)-CSIC, 28049 Madrid, Spain*

(Dated: November 5th, 2010)

---
[a)] Invited speaker and author for correspondence: xavierbatlle@ub.edu

**Abstract.** The magnetic behavior of $Fe_{3-x}O_4$ nanoparticles synthesized either by high-temperature decomposition of an organic iron precursor or low-temperature co-precipitation in aqueous conditions, is compared. Transmission electron microscopy, X-ray absorption spectroscopy, X-ray magnetic circular dichroism and magnetization measurements show that nanoparticles synthesized by thermal decomposition display high crystal quality and bulk-like magnetic and electronic properties, while nanoparticles synthesized by co-precipitation show much poorer crystallinity and particle-like phenomenology, including reduced magnetization, high closure fields and shifted hysteresis loops. The key role of the crystal quality is thus suggested since particle-like behavior for particles larger than about 5 nm is only observed when they are structurally defective. These conclusions are supported by Monte Carlo simulations. It is also shown that thermal decomposition is capable of producing nanoparticles that, after further stabilization in physiological conditions, are suitable for biomedical applications such as magnetic resonance imaging or bio-distribution studies.

**INTRODUCTION**

Nanostructured magnetic materials comprise a very active research field due to the new phenomena taking place at the nanoscale as a consequence of the interplay of quantum, finite-size, surface and interfacial effects. Magnetic nanoparticles (NPs) [1] are an excellent example of nanostructured materials, which provide the critical building blocks for the booming of many applications of nanotechnology in fields such as biomedicine [2], high-density magnetic recording [3] or magnetic resonance imaging (MRI) [4]. A key question in these systems is how the nanostructure modifies their magnetic and electronic properties and how one can take advantage of those new properties to improve applications. Consequently, understanding and controlling the effects of the nanostructure on the properties of the particles have become increasingly relevant issues for technological applications.

As the size of particles is reduced below about 100 nm, deviations from bulk behavior have been widely reported and attributed to changes in the magnetic ordering at the surface layer with respect to that of the particle core and also to finite-size effects, thus giving rise to a significant decrease in the particle magnetization [1]. Typically, transition metal oxide NPs, including iron oxide, show surface spin disorder due to symmetry breaking at the surface, leading to [5-8]: (i) saturation magnetization smaller than in the bulk material, in some cases by a factor of two or more; (ii) high-field differential susceptibility; (iii) extremely high closure fields (even up to 120 kOe); (iv) shifted hysteresis loops after field cooling the sample together with glassy behavior, which have been explained in terms of the existence of a surface layer of disordered spins that freeze in a spin glass-like state due to magnetic frustration, yielding both an exchange field acting on the ordered core and an increase in the particle anisotropy; and (v) extremely high magnetic anisotropy associated with the fact that very high magnetic fields are required to reach magnetic saturation.

However, that particle-like behavior is undesirable in many applications, where the NPs should comply with a variety of requirements. In the case of biomedical applications, those comprise [2]: (i) superparamagnetic behavior at room temperature in order to avoid particle aggregation due to inter-particle interactions; (ii) high saturation magnetization to achieve a high magnetic response under the application of moderate magnetic fields; (iii) limiting size for *in vivo* applications within 1 and 50 nm; and (iv) bio-compatibility and functionality to reach specific targets inside the body. Consequently, biomedical applications require of magnetic NPs of a few nanometers in size showing magnetic properties close to the bulk material. Research in this field is thus mainly focused on new synthesis methods enabling the preparation of NPs with high crystal quality and reduced magnetic disorder. Iron oxide NPs are suitable candidates due to their low toxicity and ease to be functionalized. Magnetite $Fe_3O_4$ and maghemite $\gamma$-$Fe_2O_3$ are probably among the most widely studied magnetic materials. However, they still arise much attention since several questions remain open. For example, the orbital contribution to the magnetic moment in bulk magnetite is under discussion [9].

In this paper, we report on the effect of the synthesis method and surface coating on the structural quality and magnetic properties of two sets of samples of $Fe_{3-x}O_4$ NPs of about 5 nm, synthesized either by (i) thermal decomposition of an organic iron precursor in an organic medium [10], with a fatty acid (e.g., oleic acid) used as surfactant covalently bonded to the particle surface [11]; or (ii) co-precipitation of Fe[II] and Fe[III] salts in an alkaline aqueous medium [12], with polyvinyl alcohol (PVA) used as a protective coating against oxidation adsorbed at the NP surface but without any chemical bond. The crystal quality of both sets of samples determines the occurrence of particle-like phenomenology, which cannot be considered as intrinsic properties arising from finite-size and surface effects.

**SAMPLE PREPARATION**

Oleic acid-coated particles of 5 nm in mean diameter were synthesized by thermal decomposition of an iron precursor [10,11]. One mmol of Fe(III)-acetylacetonate was mixed with 3 mmol of oleic acid, 3 mmol of oleilamine, 5 mmol of 1,2-hexadecanediol and 10 ml of phenyl ether. The degassed mixture was heated to 200 °C and kept at that temperature for two hours. Then, it was further heated up to the reflux temperature of phenyl ether (260~270 °C) and kept at that temperature for 30 minutes. After cooling down to room temperature, the particles were extracted by solvent-non solvent techniques and finally re-suspended in hexane. Interestingly enough, variations of this procedure enable to tune the size and shape of the particles. For example, solvents with higher boiling point lead to larger particles [10] or the use of decanoic acid instead of oleic acid as capping ligand yields particles in a wider range of sizes [13]. By varying the molar ratio of the decanoic acid to the iron precursor and adjusting the final synthesis temperature, NPs within 5 and 50 nm are obtained [13]. The particle size range can be further expanded up to 180 nm by tuning the heating rate [14].

PVA-coated 5 nm in size NPs were prepared by the standard co-precipitation procedure [12]. A stoichiometric mixture of $FeCl_3$ and $FeCl_2$ was dissolved in 10 ml de-ionized water and added drop-wise to 300 ml of a sodium hydroxide aqueous solution, containing PVA at room temperature and pH 10. The particles formed instantaneously, were extracted by precipitation and were finally suspended in de-ionized water at pH 7.

**STRUCTURAL CHARACTERIZATION**

High resolution transmission electron microscopy (TEM) images of NPs from thermal decomposition show regular shapes with very high crystal quality up to the NP surface (Figs.1 (a)-(b)). In contrast, NPs from co-precipitation show much more irregular shapes and lower degree of crystallinity at the surface (Figs.1 (c)-(d)), together with in-volume defects. Many of those particles are aggregates of individual crystallites randomly oriented. TEM also shows that the particles prepared by thermal decomposition have a narrow size distribution and are individually coated by the surfactant, while those by co-precipitation tend to agglomerate and show a much broader size distribution.

Figure 2 shows an example of X-ray absorption spectroscopy (XAS) data at the Fe $L_{2,3}$ absorption edges, for $Fe_{3-x}O_4$ NPs of about 5 nm in size. The spectrum for thermal decomposition particles shows 30% more holes in the 3d band [15] as compared to the bulk counterpart [9,16], indicating an electronic transfer associated with the chemical bonding of the oleic acid at the particle surface [15]. In contrast, co-precipitation particles of similar size show a XAS spectrum similar to the bulk counterpart, as expected for particles with just a protective adsorbed coating. Both spectra can be superimposed almost perfectly, provided that the spectrum for co-precipitation be multiplied by an appropriate scale factor (inset to Fig. 2), which evidences that the stoichiometry is the same within the experimental error. The actual stoichiometry of the 5 nm NPs from thermal decomposition was evaluated from Mössbauer spectroscopy to be $Fe_{2.93}O_4$ [17], which is compatible with the presence of up to 21% of maghemite, forming an overoxidized surface layer of about one unit cell.

**PARTICLE-LIKE BEHAVIOR VERSUS STRUCTURAL QUALITY**

Magnetization hysteresis loops and zero field cooling-field cooling (ZFC-FC) curves were measured with a SQUID magnetometer. The ZFC-FC curves in Fig. 3 are compared for particles of about 5 nm in size prepared by the two methods. For thermal decomposition, the ZFC curve evidences a very narrow size distribution. The progressive increase of the FC curve below the irreversibility onset indicates the absence of relevant dipolar interactions, which is in agreement with the individual coating of the particles. The coincidence of the onset of the irreversibility and the maximum of the ZFC also confirms that the surfactant coating minimizes the formation of aggregates, avoiding inter-particle interactions. In contrast, the ZFC-FC curves for the co-precipitation sample reflects a broader size distribution and particle agglomeration, and the existence of significant inter-particle interactions. The mean activation magnetic volume of the two samples was evaluated by fitting the ZFC curves to a distribution of Langevin functions [18]. The diameter corresponding to the mean activation magnetic volume is $\langle d \rangle = 5.0$ nm in the case of thermal decomposition, with a standard deviation $\sigma = 4.4$ nm, in very good agreement with the size distribution obtained from TEM. For the co-precipitation sample, $\langle d \rangle = 27$ nm and $\sigma = 26$ nm, which are much larger than those deduced from the TEM size distribution due to particle aggregation favoring inter-particle interactions.

Figure 4 shows the hysteresis loops at 5 K for particles of about 5 nm. Saturation magnetization is significantly higher for the thermal decomposition sample. Additionally, a higher value of the high-field differential susceptibility and a much rounded magnetization curve are observed for the co-precipitation sample, facts which are indicative of the presence of magnetic disorder throughout the particles. Moreover, the magnification of the hysteresis curves (inset to Fig. 4) shows the occurrence of much higher closure fields (indicated by arrows in Fig. 4) in the co-precipitation sample, which could be related to the existence of very high energy barriers associated with magnetic frustration induced by structural disorder.

Magnetization values at 5 T and 5 K as a function of the particle size are shown in Fig. 5 [13,15,19,20]. The values corresponding to the thermal decomposition samples are always significantly higher than those for co-precipitation, both sets of data displaying some size dependence. Finite-size and surface effects are evident in the results from Monte Carlo simulations, also shown in Fig. 5 (solid triangles) for a spherical particle with Ising spins on an inverse spinel lattice [21]. These data have been scaled to the saturation magnetization of bulk magnetite in order to be compared to the experimental results. Experimental values lay clearly below the ones obtained by Monte Carlo simulation, even for thermal decomposition samples. This could be due to the difficulty in obtaining pure, stoichiometric magnetite without a significant amount of the maghemite phase for particle sizes below a few tens of nm. In the case of co-precipitation, the lack of crystal quality yields a large additional reduction in the magnetization due to magnetic disorder within the particle.

X-ray magnetic circular dicroism (XMCD) was measured at the Fe $L_{2,3}$ absorption edges for $Fe_{3-x}O_4$ NPs of about 5nm in size [15]. The measurements were carried out with total electron yield detection and an applied field of 2 T parallel to the X-ray beam. The application of magneto-optical sum rules on the XMCD spectra enables to obtain the magnetic moment per formula unit, $\mu_{Fe}$, and the spin and orbital contributions [22] (Table I). The total magnetic moment per formula unit for the thermal decomposition sample is about 42% larger than that for the co-precipitation sample, in excellent agreement with magnetization results. Furthermore, $\mu_{Fe}$ for the thermal decomposition sample is just about 16% smaller than that reported from previous XMCD measurements on a magnetite single-crystal [9], also in excellent agreement with magnetization results. Besides, the calculated orbital magnetic moments are 0.036(8) and 0.081(9) $\mu_B$/f.u. for the thermal decomposition and co-precipitation samples, respectively. Although both values are in qualitative agreement with Reference 9, the co-precipitation sample shows about a 3 fold increase with respect to the expected value of the ratio of the orbital-to-spin angular moments $m_L/m_S \approx 0.013$, obtained from LDA calculations for bulk magnetite [23]. Increased values of $m_L/m_S$ are usually found in low dimensional and particle-like systems [24]. In contrast, the thermal decomposition sample fully recovers the value for the bulk orbital-to-spin moment, which reinforces the idea that the high crystal quality of the sample is behind the observed bulk-like magnetic and electronic behavior [15].

In order to elucidate the influence of the crystal quality and particle/surfactant chemical bond of thermal decomposition samples on the magnetic disorder at the NP surface, we evaluated the surface and volume anisotropies ($K_s$ and $K_v$, respectively) from the size distribution obtained from TEM and $T \ln(t/\tau_0)$ scaling of the thermo-remanence relaxation data [7,8,25,26]. In the absence of sizeable inter-particle interactions, as is the case for the thermal decomposition particles (see Fig. 3), the derivative of the $T \ln(t/\tau_0)$ scaling yields the energy barrier distribution associated with the effective anisotropy for magnetization reversal. Assuming the *ad hoc* expression for the effective, uniaxial anisotropy, $K_{eff} = K_v + 6/D K_s$, the size distribution obtained from TEM can be superimposed onto the energy barrier distribution by a fitting procedure that yields the volume and surface anisotropies [27]. The values obtained are given in Table II, being in very good agreement with other experimental values (see 28, 27 and references therein). Thus, intrinsic surface effects, rather than the lack of crystallinity, dominate the surface anisotropy in high crystal quality particles nanometer in size.

Another effect usually related to magnetic disordered systems is the occurrence of shifted magnetization hysteresis loops, which in many cases is explained assuming the existence of the exchange bias phenomenon [29,30]. Shifted loops can be observed after FC processes, or even ZFC, for samples prepared by co-precipitation [19,31], as shown in Fig. 6 for 5 nm $Fe_{3-x}O_4$ PVA-coated NPs, having poor crystallinity. Additionally, maghemite hollow particles of 8 nm in size composed of about 10 crystallites randomly oriented, show strongly shifted loops and very high closure fields due to their highly defective crystallographic structure and the large interface areas among single crystallites [32]. All those systems have in common the occurrence of magnetic disorder associated with defective crystal structures, giving rise to high energy barriers that can be only overcome at very high fields. On the contrary, thermal decomposition samples with high crystal quality do not show this phenomenology at all. For very defective particles showing high closure fields (inset to Fig. 4), the applied magnetic field of a FC process imprints an effective anisotropy that cannot be erased at the maximum applied field of the hysteresis loop. Therefore, the system follows a minor loop that, in some cases, has been attributed to an exchange coupling phenomenon between a core-shell structure [29,31,33].

In order to reinforce this interpretation, Fig. 7 shows two hysteresis loops obtained by atomistic Monte Carlo simulations of a single spherical maghemite particle of 4 unit cells in diameter, with 40% of vacancies in the Fe positions at the surface [21]. Both magnetization curves were obtained after FC down to 0.5 K in a field of $h_{FC} = 20$ K (the field is given in temperature units as $\mu H / k_B$), while the maximum applied field was 80 K in one case and 200 K in the other. It is obvious that the apparent shifted loop in the case of $h = 80$ K is just a minor loop of that corresponding to h = 200 K, because it was obtained for a maximum field lower than the closure field (approximately 90 K). Similar apparent shifted loops, which in fact are minor loops, have also been reported [32].

**MAGNETITE NANOPARTICLES FOR BIOMEDICAL APPLICATIONS**

As iron oxides are approved for biomedical use in humans due to their low toxicity, magnetite is the material of choice when the highest signal is desired, since its saturation magnetization per unit volume is about 14% higher than that of maghemite. We have shown that thermal decomposition of an organic Fe precursor is an appropriate method to obtain magnetite NPs with high structural and magnetic quality. However, this chemical route yields hydrophobic particles that require of further functionalization and stabilization under physiological conditions. This implies surface modification by attaching specific ligands and/or coating with inorganic shells. As noted above, the use of decanoic acid as capping ligand yields particles of regular shapes, with high crystal quality up to the surface in a wide range of sizes [13]. The relatively short hydrocarbon chain of decanoic acid makes these NPs dispersible in slightly polar organic media, such as alcohols. This is an advantage when further coating of the particles is required, since usually that reactions take place in polar solvents, as is the case of $SiO_2$ coating by the Stöber method [34].

On the other hand, when oleic acid is used as capping ligand, it is possible to ligand exchange the NPs to dimercaptosuccinic acid (DMSA), which renders the particle surface hydrophilic [35]. Although the ligand exchange process reduces the saturation magnetization in about 10% due to some surface oxidation and promotes the formation of a small amount of aggregates constituted of a few particles, the magnetic quality is still much higher than that of the NPs prepared by co-precipitation [36]. These DMSA-coated NPs display MRI relaxativity values higher than those of commercially available contrast agents. Moreover, it also is possible to obtain high-contrast MRI images of liver and brain of mice injecting the particles intravenously and making them overcome the brain blood barrier by osmotic disruption [36]. The high magnetic signal of these particles enabled us to carry out a bio-distribution study by means of magnetization measurements, which could be an excellent alternative to the usual histograms obtained from microscopy. The magnetization for samples of lyophilized liver, kidney and spleen of mice was measured, showing the presence of diamagnetic material (tissue), paramagnetic natural ferritin, and superparamagnetic particles. The comparison of the signal of the samples extracted from inoculated animals to those of control animals allowed determining accurately the percentage of internalization of the NPs with respect to the total dose, leading to 35% for liver, 1.5% for spleen, and 0.5% for kidney [36].

**SUMARY AND CONCLUSIONS**

We have shown that in magnetic NPs, bulk-like behavior is strongly tied to the crystal quality of the samples. When particles are larger than about 5 nm in size, particle-like behavior, including strongly reduced magnetization, high values of the closure fields and shifted loops, are only observable in NPs with poor crystallinity. This is supported by Monte Carlo simulation showing that intrinsic finite-size and surface effects [1] are relevant only for sizes below about 5 nm. We have also shown that thermal decomposition is a very versatile method to produce high quality nanoparticles with bulk-like properties within 5 and 50 nm in size. These particles are also suitable for biomedical applications. All in all, these results suggest the key role of the crystal quality on the magnetic and electronic properties of ferrimagnetic NPs and, in particular, the fact that, in many cases, the magnetic disorder phenomena observed in single-phase particles larger than a few nanometers in diameter should not be considered as an intrinsic effect associated with the finite size.

**ACKNOWLEDGEMENTS**

The funding of the Spanish MICINN (MAT2009-08667, MAT2008-01077 and CSD2006-00012), Catalan DIUE (2009SGR856) and Aragonese DGA (CAMRADS and IMANA) is acknowledged.

**TABLES**

| Sample | $N_h$ | $\mu_L$ ($\mu_B$) | $\mu_S$ ($\mu_B$) | $\mu_{Fe}$ ($\mu_B$) | $m_L/m_S$ (%) |
|---|---|---|---|---|---|
| Thermal decomposition | 17.5 (3) | 0.036 (8) | 3.24(5) | 3.27(6) | 1.1(3) |
| Co-precipitation | 13.7 (3) | 0.081 (9) | 2.25(3) | 2.31(3) | 3.6(4) |

**Table I.** Total number of 3d holes, $N_h$, magnetic moment per formula unit, $\mu_L$ (orbital), $\mu_S$ (spin) and $\mu_{Fe}$ (total), and orbital-to-spin moment ratio, $m_L/m_S$, for oleic acid and PVA samples.

| | Fitted | Expected |
|---|---|---|
| $K_v$ | $2.4 \times 10^5$ erg cm$^3$ | $2.2 \times 10^5$ erg cm$^3$ (below the Verwey transition) |
| $K_s$ | 0.029 erg cm$^{-2}$ | 0.02-0.04 erg cm$^{-2}$ |

**Table II.** Fitted values of the volume and surface anisotropy constants, $K_v$ and $K_s$ respectively, for thermal decomposition samples.

**FIGURE CAPTIONS**

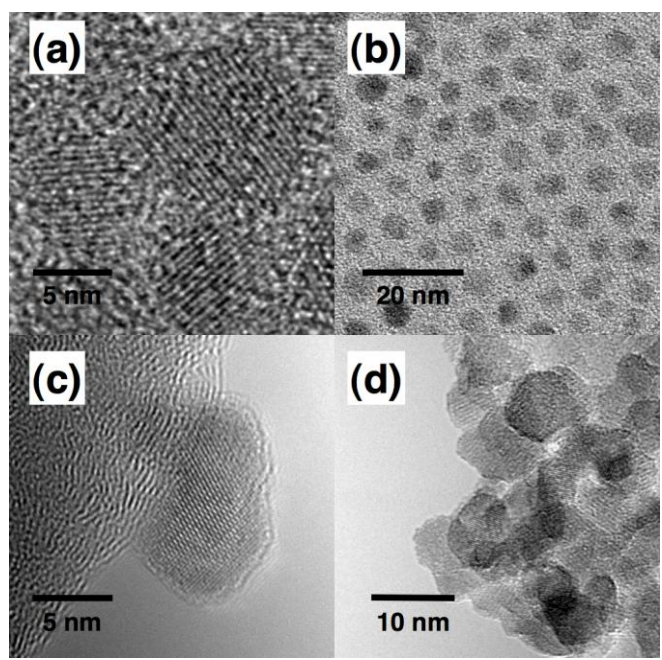

**Figure 1**. Oleic acid-coated $Fe_{3-x}O_4$ NPs prepared by thermal decomposition: [(a) and (b)] high resolution TEM images, showing high crystal quality and individual coating by the surfactant. PVA-protected $Fe_{3-x}O_4$ NPs prepared by co-precipitation: [(c) and (d)] high resolution TEM images, showing poorer crystallinity as compared to the previous case and particle aggregation.

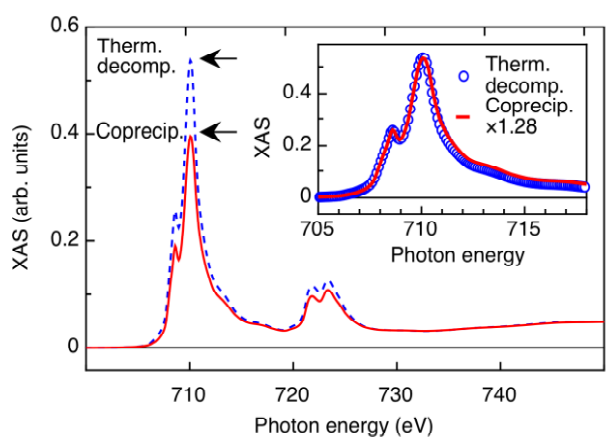

**Figure 2**. X-ray absorption spectra in the Fe $L_{2,3}$ edges for $Fe_{3-x}O_4$ NPs of about 5 nm in size, synthesized by thermal decomposition (dashed curve) and co-precipitation (solid curve).

Inset: Superimposition of the spectra for the two samples, by multiplying the spectrum of the co-precipitation sample by an appropriate scale factor.

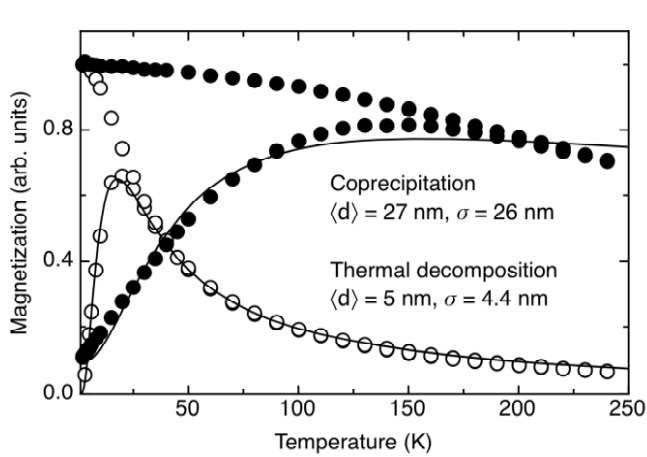

**Figure 3**. ZFC-FC curves measured at 50 Oe for $Fe_{3-x}O_4$ NPs of about 5 nm in size, obtained by thermal decomposition (open circles) and co-precipitation (solid circles). The fits to a distribution of Langevin functions are plotted as solid lines. The diameter *<d>* corresponding to the mean activation magnetic volume and standard deviation σ are indicated.

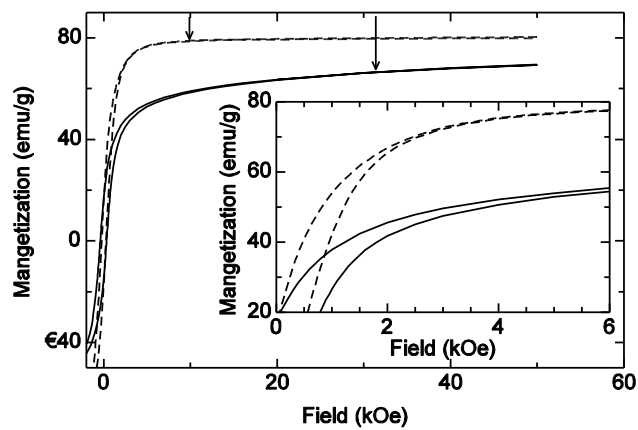

**Figure 4**. Magnetization curves at 5 K for $Fe_{3-x}O_4$ NPs of about 5 nm in size obtained by thermal decomposition (dashed lines) and co-precipitation (solid lines). The closure of the hysteresis loops are indicated by vertical arrows. Inset: detail of the curves for magnetic fields below 6 kOe.

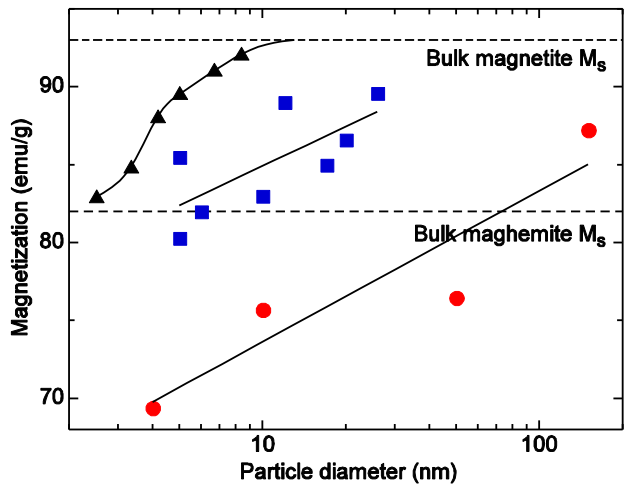

**Figure 5**. Magnetization as a function of the particle diameter for Fe$_{3-x}$O$_4$ NPs synthesized by co-precipitation (solid circles) and thermal decomposition (solid squares), at 5 T and 5 K. Spontaneous magnetization obtained from Monte Carlo simulations, scaled to the saturation magnetization of bulk magnetite (solid triangles). Saturation magnetization $M_s$ for bulk magnetite and maghemite are indicated as dashed horizontal lines. Solid lines are guides for the eye.

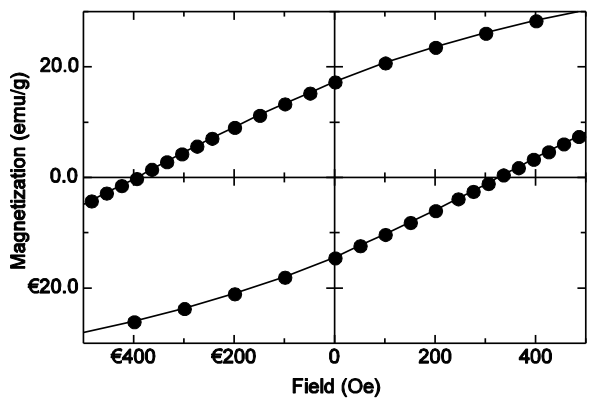

**Figure 6**. Detail of the hysteresis loop at 5 K for 5 nm Fe$_{3-x}$O$_4$ PVA-coated NPs synthesized by co-precipitation, measured after field cooling the sample at 2 kOe, showing loop shift to negative fields.

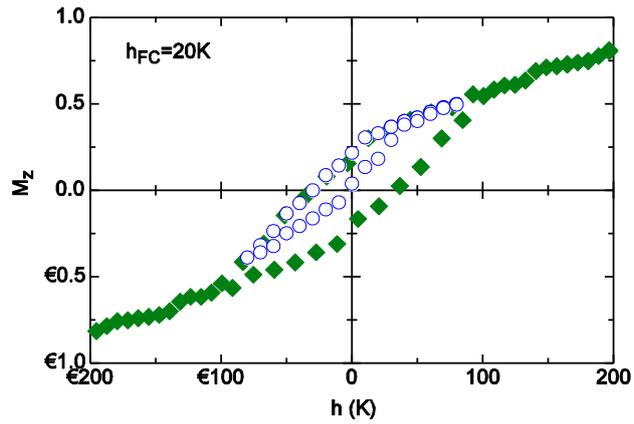

**Figure 7**. Atomistic, Monte Carlo simulation of two magnetization, $M_z$, hysteresis cycles at 0.5 K of a spherical maghemite particle of 4 unit cells in diameter with 40% vacancies at surface Fe positions. The particle was field cooled (FC) at a reduced field of 20 K and the maximum field applied in the cycles was either 80 K (open circles) or 200 K (solid squares).